\documentclass{appolb}
\usepackage{array}
\usepackage{color}
\usepackage{graphicx}

\newcommand{\tpz}{${}^{3\!}P_0$}

\newcommand{\kbw}{k_{\mbox{\scriptsize BW}}}
\newcommand{\kp}{k_{\mbox{\scriptsize pole}}}
\newcommand{\kmax}{k_{\mbox{\scriptsize max}}}

\newcommand{\ep}{E_{\mbox{\scriptsize pole}}}
\newcommand{\mbw}{M_{\mbox{\scriptsize BW}}}
\renewcommand{\mp}{M_{\mbox{\scriptsize pole}}}
\newcommand{\gbw}{\Gamma_{\mbox{\scriptsize BW}}}
\newcommand{\gp}{\Gamma_{\!\mbox{\scriptsize pole}}}
\begin{document}
\title{Dramatic implications of unitarity for meson spectroscopy\thanks
{Presented by G.~Rupp at Workshop ``Excited QCD 2019'',
Schladming, Austria, Jan.\ 30 -- Feb.\ 3, 2019.}}
\author{George Rupp\address{Centro de F\'{\i}sica e Engenharia de Materiais
Avan\c{c}ados, Instituto Superior T\'{e}cnico, Universidade de Lisboa,
P-1049-001, Portugal}
\\[2mm]
Eef van Beveren\address{Centro de F\'{\i}sica da UC, Departamento de
F\'{\i}sica, Universidade de Coimbra, P-3004-516, Portugal}
}
\maketitle
\begin{abstract}
An unambiguous definition of meson resonance masses requires a description
of the associated phase shifts in terms of a manifestly unitary $S$-matrix
and its complex poles. However, the commonly used Breit-Wigner
(BW) parametrisations can lead to appreciable deviations. We demonstrate
this for a simple elastic resonance, viz.\ $\rho(770)$, whose pole and BW
masses turn out to differ by almost 5 MeV. In the case of the very broad
$f_0(500)$ and $K_0^\star(700)$ scalar mesons, the discrepancies are shown to
become much larger, while also putting question marks at the listed PDG BW
masses and widths. Furthermore, some results are reviewed of a manifestly
unitary model for meson spectroscopy, which highlight the potentially huge
deviations from static model predictions. Finally, a related unitary model
for production amplitudes is shown to explain several meson enhancements
as non-resonant threshold effects, with profound implications for
spectroscopy.
\end{abstract}
\section{Introduction}
The most fundamental cornerstone of the PDG tables is the uniqueness of
$S$-matrix pole positions of unstable particles, as a consequence of
quantum-field-theory principles. Therefore, the unitarity property of the
$S$-matrix should ideally be respected in whatever description of mesonic
resonances in experiment, on the lattice, and in quark models. However,
simple Breit-Wigner (BW) parametrisations that not always
satisfy unitarity continue to be widely used in data analyses of mesonic
processes. In this short paper, the resulting discrepancies will be studied
for three elastic meson resonances, viz.\ $\rho(770)$, $f_0(500)$ (alias
$\sigma$), and $K_0^\star(700)$ (alias $\kappa$).

Now, quark models usually treat mesons as 
permanently bound $q\bar{q}$ states, ignoring the dynamical effects of
strong decay, be it real or virtual. Only a model that respects $S$-matrix
unitarity of the decay products can be reliably compared to resonances in
experiment. A few important results of models employed by us since long
ago will be reviewed here. Finally, some predictions of a strongly related
unitary model of productions processes, with far-reaching consequences for
meson spectroscopy, will be briefly revisited. \\[-4mm]
\section{Pole mass vs.\ Breit-Wigner mass}
Now we summarise very succinctly how to relate the pole mass of an elastic
resonance to its typical Breit-Wigner (BW) mass, with some applications.
A detailed derivation will be published elsewhere.

A $1\times1$ partial-wave $S$-matrix, being a function of the relative
momentum $k$, can be written as \cite{Taylor} $S_l(k)=J_l(-k)/J_l(k)$,
where $J_l(k)$ is the so-called Jost function. A resonance then corresponds
to a simple pole in $S_l(k)$ for complex $k$ with positive real part
and negative imaginary part, that is, a pole lying in the fourth
quadrant of the complex $k$-plane. So the simplest ansatz for the $S$-matrix
and thus for the Jost function is to write $J_l(k)=k-\kp = k-(c-id)$,
with $c>0 , d>0$.
Note that this requires $S_l(k)$ to have a zero in the
second quadrant, viz.\ for $k=-c+id$.
But then the $S$-matrix cannot be unitary \cite{Kaminski}, for real $k$, i.e.,
$S_l^\star(k)\neq S_l^{-1}(k)$. It will only satisfy unitarity if
\cite{Taylor} the Jost function obeys $J_l^\star(k)=J_l(-k)$, for real $k$.
Consequently, the Jost function should read \cite{Kaminski}
$J_l(k)=(k-\kp)(k+\kp^\star)=(k-c+id)(k+c+id)$.
So $S_l(k)$ has a symmetric pair of poles in the 3rd and 4th quadrants,
corresponding to an equally  symmetric pair of zeros in the 1st and 2nd
quadrants. Note that in the complex-energy plane, given by
$E=2\sqrt{k^2+m^2}$ in the case of two equal-mass particles, this amounts to
one pole and one zero lying symmetrically in the 4th and 1st quadrants,
respectively. Since a $1\times1$ $S$-matrix can generally be written as
$S_l(k)=\exp(2i\delta_l(k))=(1+i\tan\delta_l(k))/(1-i\tan\delta_l(k)$),
we can use the unitary expression for the Jost function above to derive
\cite{Kaminski} \mbox{ } \\[-2mm]
\begin{equation}
\tan\delta_l(k) \; = \; \frac{2k\,\mbox{Im}(\kp)}{k^2-|\kp|^2} \; = \;
\frac{2dk}{c^2+d^2-k^2} \; . \\[-1.5mm]
\label{tandelta}
\end{equation}
When the partial-wave phase shift $\delta_l(k)$ reaches $90^\circ$,
we get for the modulus of the corresponding amplitude
\mbox{$|T_l(\kmax)|\!=\!|\exp(i\delta_l(\kmax))\sin\delta_l(\kmax)|=1$},
for $\kmax^2=c^2+d^2$. The associated maximum energy
$E_{\mbox{\scriptsize max}}=2\sqrt{\kmax^2+m^2}=2\sqrt{c^2+d^2+m^2}$
is different from the maximum in a typical Breit-Wigner (BW) amplitude
$T_l(E)\propto (E-\mbw+i\gbw/2)^{-1}$, which is called the BW mass and just
given by the real part of the pole in the fourth
quadrant of the complex-energy plane, viz.\
$\mbw=2\sqrt{\kbw^2+m^2}=2\sqrt{c^2+m^2}$. Such a BW amplitude, in 
spite of being unitary in the case of an isolated resonance, can give rise to
significant differences compared to $S$-matrix approaches.

Next we illustrate the consequences of these unitarity considerations in the
simple case of the very well-known meson $\rho(770)$ \cite{PDG2018}, which
is an elastic $P$-wave resonance in $\pi\pi$ scattering. The PDG 
lists its mass and total width as \cite{PDG2018} $M_{\rho^0}=(775.26\pm0.25)$
MeV and $\Gamma_{\!\!\rho^0}=(147.8\pm0.9)$ MeV,
where the width follows almost exclusively ($\approx100$\%) from the decay mode
$\rho^0\to\pi^+\pi^-$, with $m_{\pi^\pm}=139.57$~MeV.

In the following, we shall refer to BW mass ($\mbw$) for the energy where
the resonance's phase shift passes through $90^\circ$ and so the modulus of
the amplitude is maximum. This also holds for the standard 
BW amplitude given above, though in the latter case it corresponds to the real
part of the resonance pole's complex energy. In contrast, here we want to
determine the difference between pole mass and (unitary) BW mass for the
$\rho(770)$. After some lengthy yet straightforward algebra, we can express
the pole mass explicitly in terms of the BW mass and the pole width as \\[-1mm]
\begin{equation}
\mp \; = \; \sqrt{\sqrt{(\mbw^2-4m^2)^2-4m^2\,\gp^2}+4m^2-\gp^2/4} \; .
\mbox{ } \\[-1mm]
\label{mp}
\end{equation}
Note that it is not possible to write $\mp$ as a simple closed-form expression
in terms of both $\mbw$ and $\gbw$. Assuming for the moment that
$\gp\simeq\gbw$, we substitute in Eq.~(\ref{mp}) the PDG values for
$\mbw$ and $\gp$, which yields $\mp=770.67$ MeV. This is 4.5~MeV lower
than the PDG $\rho(770)$ mass of 775.25~MeV!
Now we check whether indeed $\gp\simeq\gbw$, by calculating the
half-width of the $\rho(770)$ peak from the modulus squared of the amplitude
$T_l(k)$, starting from Eq.~(\ref{tandelta}). The result is $\gbw=147.83$~MeV,
so indeed very close to the assumed $\gp=147.8$~MeV.
Finally, we compare pole mass and width vs.\ BW mass and width for the 
very broad scalar mesons $f_0(500)$ and $K_0^\star(700)$ \cite{PDG2018}. As
the latter resonance decays into $K\pi$, we must now deal with the unequal-mass
case, which does not allow to derive simple expressions. Yet on the
computer the real and imaginary parts of $\kp$ can be simply obtained,
allowing to derive $\mbw$ and $\gbw$ as before.

Let us now check what the consequences are for $f_0(500)$
and $K_0^\star(700)$. Their pole positions as well as BW masses and widths are
listed in the PDG Meson Tables as \cite{PDG2018}
\begin{eqnarray}{l}
\hspace*{-5mm}
f_0(500): \;\; 
\left\{\begin{array}{l}
\ep\;=\;\left\{(475\pm75)-i(275\pm75)\right\} \mbox{MeV}\;,
\\[1mm]
\mbw \; = \; (475 \pm 75) \; \mbox{MeV} \;,\;\;
\gbw \; = \; (550 \pm 150) \; \mbox{MeV} \; ;
\end{array}\right. \\
\label{sigmapdg}
K_0^\star(700): \;\;
\left\{\begin{array}{l}
\ep\;=\;\left\{(680\pm50)-i(300\pm40)\right\} \mbox{MeV}\;,
\\[1mm]
\mbw \; = \; (824 \pm 30) \; \mbox{MeV} \;,\;\;
\gbw \; = \; (478 \pm 50) \; \mbox{MeV} \; .
\end{array}\right.
\label{kappapdg}
\end{eqnarray}
But using our equations imposed by elastic $S$-matrix
unitarity, we obtain \\[-2mm]
\begin{equation}
\begin{array}{rl}
f_0(500): & \!\!\mbw \; = \; (592 \pm 99) \; \mbox{MeV} \; , \;\;
\gbw \; = \; (555 \pm 152) \; \mbox{MeV} \; ; \\[1mm]
K_0^\star(700): & \!\!\mbw \; = \; (907 \pm 49) \; \mbox{MeV} \; , \;\;
\gbw \; = \; (709 \pm 122) \; \mbox{MeV} \; .
\label{skunitary}
\end{array}
\end{equation}
The conclusion is that the PDG seems to underestimate the BW masses of both
$f_0(500)$ and $K_0^\star(700)$, as well as the BW width of
$K_0^\star(700)$. We stress again that here ``BW'' refers
to the energy at which $\delta_l(E)=90^\circ$, in the context
of the present simple pole model. Note that reality is more complicated,
since the $f_0(500)$ resonance overlaps with $f_0(980)$ \cite{PDG2018} and
$K_0^\star(700)$ with $K_0^\star(1430)$ \cite{PDG2018}, besides the influence
of Adler zeros on the amplitudes \cite{PLB572p1}. Nevertheless, the need for
a uniform and unitary treatment of especially broad resonances in experimental
analyses is undeniable.

To conclude this section, we note that calculating $\mbw$ for $f_0(500)$
and $K_0^\star(700)$ via the cross section instead of the amplitude's modulus
becomes already problematic, while no $\gbw$ can even be defined at all. Also,
for inelastic resonances the mass discrepancy due to the use of a non-unitary
parametrisation can become as large as 170~MeV in the case of $\rho(1450)$
\cite{PRD96p113004}.  \\[-6mm]

\section{Unitarity distortions of $q\bar{q}$ spectra}
Fully accounting for unitarity when describing meson resonances, or just
computing mass shifts of $q\bar{q}$ states from real and virtual meson
loops, can give rise to enormous distortions of confinement spectra
\cite{CPC41p053104}. Moreover, it can even lead to the dynamical
generation of additional states. This allowed the unitarised
multichannel quark model of Ref.~\cite{ZPC30p615} to predict for the
first time a complete nonet of light scalar-meson resonances, whose
predicted masses and widths are still compatible with present-day PDG
limits \cite{PDG2018}. More recently, a strongly related model was
formulated \cite{IJTPGTNO11p179} in $p$-space, called
Resonance-Spectrum Expansion (RSE), resulting in a coupled-channel
$T$-matrix for non-exotic meson-meson scattering diagrammatically given by
\\[1mm]
\begin{tabular}{ccccc}
\raisebox{5mm}{$\;\;\;\;\;\;T\;\;\;=$} &
\includegraphics[trim = 15mm 0mm 0mm 0mm,clip,height=1.2cm,angle=0]
{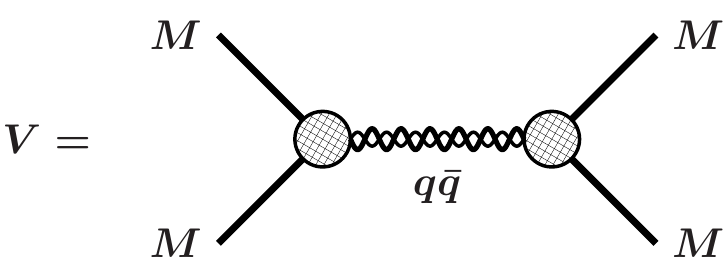} &
\raisebox{5mm}{$+$} &
\includegraphics[trim = 24mm 0mm 0mm 0mm,clip,height=1.2cm,angle=0]
{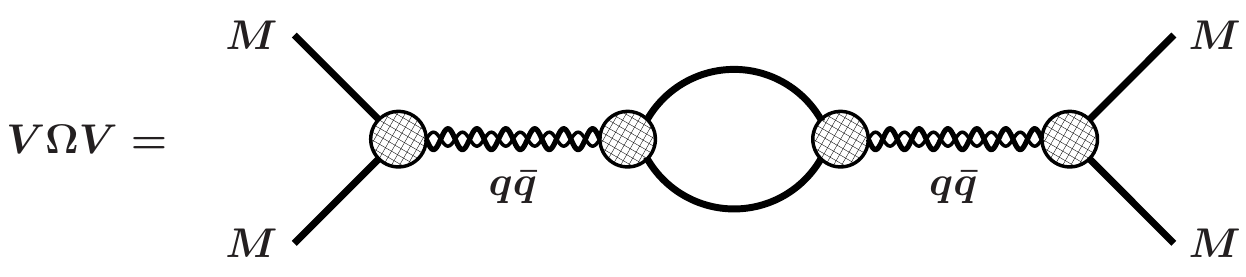} &
\raisebox{5mm}{$+\;\;\;\ldots$}
\end{tabular}
\mbox{ } \\[1mm]
Here, the wiggly lines represent a tower of bare $q\bar{q}$ states, which
couple to two-meson channels via a \tpz\ vertex. For more details and
closed-form multichannel expressions, see e.g.\ Ref.~\cite{PRD84p094020}.
Using the RSE formalism, a coupled-channel calculation of light and
intermediate scalar mesons was carried out in Ref.~\cite{APPS2p437}, 
yielding the poles \\[1.5mm]
\hspace*{1cm}$f_0(500): \;(464-i217)$~MeV, \hspace{2mm}
$f_0(1370): \;(1335-i185)$~MeV; \\[1mm]
\hspace*{1cm}$f_0(980): \;(987-i29)$~MeV, \hspace{4mm}
$f_0(1500): \;(1530-i14)$~MeV; \\[1mm]
\hspace*{1cm}$a_0(980): \;(1023-i47)$~MeV, \hspace{2mm}
$a_0(1450): \;(1420-i185)$~MeV; \\[1mm]
\hspace*{8mm}$K_0^\star(700): \;(722-i266)$~MeV, \hspace{1mm}
$K_0^*(1430): \;(1400-i96)$~MeV. \\[1.5mm]
These results are close to those found in the $r$-space
model of Ref.~\cite{ZPC30p615}. Note again the generation of two
scalar resonances for each bare $P$-wave $q\bar{q}$ state.

The possible doubling of resonances due to unitarisation becomes yet more
peculiar in cases where it is not even clear which is the intrinsic
one and which the dynamically generated state. For example, the
$D_{s0}^\star(2317)$ \cite{PDG2018} scalar $c\bar{s}$ meson showed up as a
dynamical state in a simple RSE model \cite{PRL91p012003} with only the $DK$
channel included, but as a strongly mass-shifted intrinsic state in the
multichannel RSE calculation of Ref.~\cite{PRL97p202001}. This cross-over is
demonstrated in more detail for the $\chi_{c1}(2P)$ \cite{PDG2018} (old
$X(3872)$) axial-vector $c\bar{c}$ state in Ref.~\cite{EPJC73p2351}, with being
an intrinsic or dynamical state depending on fine details of the model's
parameters. Clearly, this ambiguity in the quark-model assignment of
$D_{s0}^\star(2317)$ and $\chi_{c1}(2P)$, as well as of probably several other
mesons, has severe implications for spectroscopy. \\[-6mm]

\section{Non-resonant peaks from unitary production amplitudes}
Most meson resonances are nowadays not observed in meson-meson
scattering, mainly extracted from meson-proton data, but rather in production
processes, like e.g.\ $e^+e^-$ annihilation or $B$-meson decays. In these
situations no initial $q\bar{q}$ annihilation takes place, as the starting
point is already an isolated $q\bar{q}$ pair, resulting from a virtual
photon in $e^+e^-$  or as a decay product from a heavier meson
like e.g.\ $J/\psi$ or $B$. The corresponding production amplitude $P$ is
defined \cite{AOP323p1215} in the RSE formalism as a non-resonant,
lead term plus its infinite rescattering series via the above RSE
$T$-matrix, i.e., \\[1mm]
\begin{tabular}{ccc}
\hspace*{2mm}\raisebox{12.3mm}{$P\;\;=$} &
\includegraphics[trim = 27mm 200mm 13mm 40mm,clip,height=2.0cm,angle=0]
{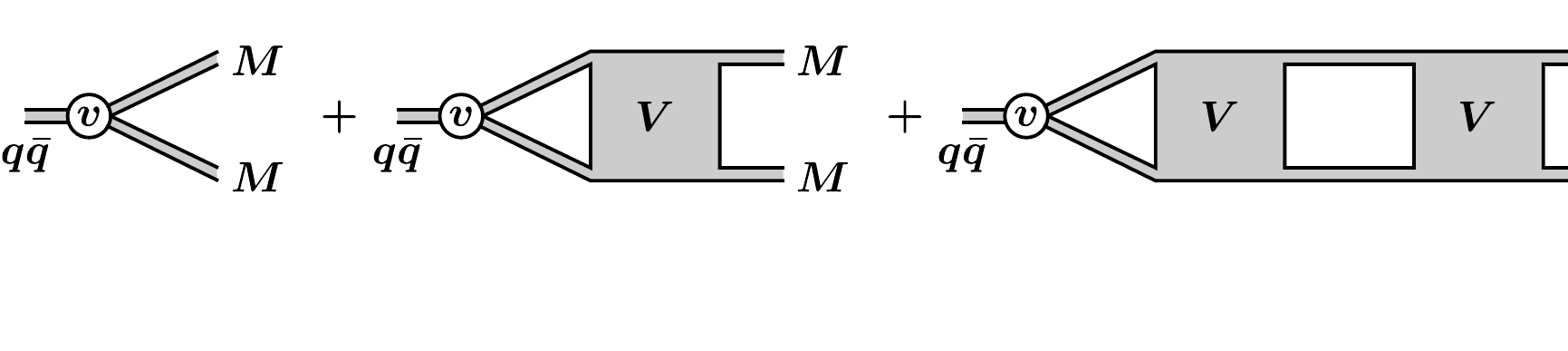}  &
\raisebox{12.3mm}{$+\;\;\ldots$}
\end{tabular}
\mbox{ } \\[-5mm]
or $P_k=\mbox{Re}(Z_k)+i\sum_l Z_l T_{kl}$,
with the $Z_k$ being purely kinematical functions related to the
$q\bar{q}$--meson-meson vertex. In the RSE model of Ref.~\cite{AOP323p1215},
where the detailed expressions can be found, the $Z_k$ are spherical
Hankel$^{(1)}$ functions and their real parts spherical Bessel functions. The
$P_k$ components satisfy \cite{AOP323p1215,EPL81p61002} the extended-unitarity
relation $\mbox{Im}(P_k)=\sum_l T^\star_{kl}\,P_l$.
Note that this can be rewritten in terms of purely imaginary
functions $\tilde{Z}_k$, so without the inhomogeneous term, but then
the real functions $i\tilde{Z}_k$ would necessarily include elements of the
$T$-matrix itself and so not be purely kinematical anymore \cite{EPL84p51002}.

There can be many applications of our production formalism in hadron
spectroscopy. In Ref.~\cite{PRD80p074001} several structures are analysed in
$K^+K^-$, $D\bar{D}$, $B\bar{B}$, and $\Lambda_c\bar{\Lambda}_c$ data. The most
dramatic conclusions are that $\Upsilon(10580)$ and $X(4660)$ (now called
$\psi(4660)$ \cite{PDG2018}) are probably not genuine resonances but rather
enhancements rising from the $B\bar{B}$ and $\Lambda_c\bar{\Lambda}_c$
thresholds, respectively.

\section{Conclusions}
We have shown unitarity to be an essential constraint in analysing
scattering data in order to allow an unambiguous determination of
resonance parameters, even in the elastic case. On the other hand, in
quark models a unitary description of meson resonances may lead to
enormous deviations from the naive bound-state spectra, and moreover give
rise to extra states not present in the bare spectra. Finally, unitarity
also plays a fundamental role in production processes, by relating them
to scattering and yielding threshold enhancements that may be mistaken
for true resonances. The consequences for modern meson spectroscopy are
far-reaching. \\[-9mm]

\end{document}